\documentclass[11pt,english,11pt,a4paper]{article}
\usepackage[T1]{fontenc}
\usepackage[latin9]{inputenc}
\usepackage{amsmath}

\makeatletter
\usepackage{jheppub}
\usepackage{slashed}
\usepackage[yyyymmdd]{datetime}
\usepackage{xcolor}

\makeatother

\usepackage{babel}
\usepackage{fancybox,xcolor}
\usepackage{empheq}

\definecolor{cardinal}{rgb}{0.77, 0.12, 0.23}

\begin{document}
\title{Dark phase transition from WIMP: complementary tests from gravitational waves and colliders}

\abstract{A dark sector is an interesting place where a strong first-order phase transition, observable gravitational waves and/or a dark matter candidate   could arise. However, the experimental tests for such a dark sector could be ambiguous due to the dark content, largely  unconstrained  parameter space and the connection  to the visible world.
We consider   a minimal dark scalar-vector boson plasma to realize the three mentioned phenomena, with a unique connection to the Standard Model via the Higgs portal coupling.  We discuss the important features of the Higgs portal  in such a minimal dark sector, namely the dark thermalization, collider tests, and direct detection of    dark matter. 
We perform     numerical analyses of  the dark phase transition associated with stochastic  gravitational waves and  dark matter, discussing the   complementarity of    collider detection,  dark matter direct detection and space-based/terrestrial   interferometers as a promising avenue to  hear and see the minimal dark sector.

}

\author[a]{Shinya Kanemura}
\author[b]{and Shao-Ping Li} 

\affiliation[a]{Department of Physics, Osaka University, Toyonaka, Osaka 560-0043, Japan} 
 \emailAdd{kanemu@het.phys.sci.osaka-u.ac.jp}
\affiliation[b]{Institute of High Energy Physics, Chinese Academy of Sciences, Beijing 100049, China} 
 \emailAdd{spli@ihep.ac.cn}
 
 \preprint{OU-HET 1199}

\maketitle

\section{Introduction}
An Abelian dark   gauge sector is one of the simplest scenarios that could undergo a strong first-order phase transition in the early universe and produce observable gravitational waves (GW)~\cite{Schwaller:2015tja}.  The dark sector is also a natural place for the  dark matter (DM) candidate which annihilates into dark species in a weakly-interacting-massive-particle (WIMP) like pattern~\cite{Pospelov:2007mp,Feng:2008mu,Berlin:2016gtr,Evans:2017kti}. The dark sector could also be detectable at colliders, depending on the portal couplings to the Standard Model (SM)~\cite{Batell:2022dpx,Ferber:2023iso}.


A strong first-order phase transition in a dark Abelian gauge  sector  was widely considered via    a   scale-invariant scalar potential~\cite{Jaeckel:2016jlh,Jinno:2016knw,Chiang:2017zbz,Brdar:2018num,Marzo:2018nov,YaserAyazi:2019caf,Mohamadnejad:2019vzg,Ellis:2019oqb,Ellis:2020nnr} or a  scale-breaking one~\cite{Addazi:2017gpt,Croon:2018erz,Breitbach:2018ddu,Hasegawa:2019amx,Dent:2022bcd,Li:2023bxy}.   
Moreover, the interesting connection among the  phase transition, GW and DM has received increased attention~\cite{Baldes:2017rcu,Huang:2017kzu,Alves:2018jsw,Baldes:2018emh,Bian:2018mkl,Mohamadnejad:2019vzg,Hong:2020est,Costa:2022oaa,Kierkla:2022odc,Costa:2022lpy,Chen:2023rrl,Arcadi:2023lwc}, where complementary tests  from DM detection and GW interferometers can be  a promising avenue to  probe the dark sector. In this respect, the hidden world becomes   not so \textit{dark} as its name suggests, since we will be able  to not only hear but also see   the dark. 

In this paper, we consider   a   dark gauge $U(1)_X$ sector to realize a strong first-order phase transition,  GW signals and a stable DM candidate. The results presented in this paper will differ from previous  work in several aspects.  First of all, unlike a scale-invariant dark sector with a radiative symmetry breaking~\cite{Coleman:1973jx}, we consider a dark phase transition with a vacuum mass term, so that the   $U(1)_X$ symmetry breaking occurs spontaneously via the Higgs-like mechanism.
On the other hand,  
  we take a minimal perspective to consider a  scale-breaking dark sector such that it only consists  of a dark scalar and a dark gauge boson (a boson plasma). Without any fermionic DM, the dark gauge boson carrying an unbroken $Z_2$ symmetry can itself play the role of a WIMP-like DM candidate. Besides,  the dark phase transition is dominantly triggered by  the finite-temperature correction from the DM,   therefore  allowing one to obtain stronger connection among the dark phase transition, GW and DM, and in particular for the   complementary probes. 

In addition to  the minimal setup of the   dark boson plasma, we further  discuss the important role of the Higgs portal coupling, which is the unique renormalizable connection between the dark boson plasma and the SM sector due to the unbroken $Z_2$ symmetry.   
We will specify three significant features of the Higgs portal for the minimal dark boson plasma, including the dark thermalization, collider detection and DM direct detection.  

Dark thermalization has been noticed in a hidden phase transition, and it was found that  if the dark sector is colder than the SM plasma~\cite{Breitbach:2018ddu}, the induced GW signals could be suppressed by a factor of $\xi^8(\ll 1)$~\cite{Li:2023bxy}, where $\xi$ is the temperature ratio of the dark and the SM sectors.  Therefore, the Higgs portal coupling is essential to induce   strong GW signals from the one-temperature ($\xi=1$) dark phase transition. The Higgs portal also opens up collider detection of the dark scalar, which acts as the mediator between the two sectors. In particular, 
direct production of a heavy dark scalar at the upcoming/proposed  high-luminosity (HL) LHC~\cite{Apollinari:2017lan}, Compact Linear Collider (CLIC)~\cite{Linssen:2012hp}, International Linear Collider (ILC)~\cite{ILC:2013jhg}, and   muon colliders~\cite{Delahaye:2019omf} could serve as the collider tests for the dark sector.  Moreover, direct detection of the gauge boson DM is also possible via a proper range of the Higgs portal which  explains the null observation of the existing DM direct detection from   LUX~\cite{LUX:2016ggv}, XENON1T~\cite{XENON:2017vdw}, PandaX-II~\cite{PandaX-II:2017hlx}, PandaX-4T~\cite{PandaX-4T:2021bab} and  LUX-ZEPLIN (LZ)~\cite{LZ:2022ufs} experiments. 

We  will  present   several benchmark points (BP) that can induce observable GW signals by space-based interferometers, including  LISA~\cite{Caprini:2019egz}, BBO~\cite{Cutler:2005qq}, and  DECIGO~\cite{Seto:2001qf},  and by some atomic interferometers, such as the km-scale terrestrial experiment AION~\cite{Badurina:2019hst} and  the strontium atomic interferometer AEDGE~\cite{AEDGE:2019nxb}. We further scrutinize each BP to seek for   complementary tests from   colliders and DM direct detection in, e.g., the LZ experiment~\cite{LZ:2015kxe,Mount:2017qzi,LZ:2018qzl} above the neutrino floor~\cite{Billard:2013qya,Ruppin:2014bra}.



The paper is outlined as follows. In Sec.~\ref{sec:framework}, we present the minimal dark $U(1)$ framework and discuss the Higgs portal interaction as the window for collider tests as well as for the dark thermalization.  In Sec.~\ref{sec:WIMPDM}, we calculate the DM relic density and discuss the DM direct detection via the Higgs portal. Then in Sec.~\ref{sec:PT}, we perform a numerical analysis of the dark phase transition, and Sec.~\ref{sec:GW} is devoted to the discussion of   GW  profiles relevant for the dark phase transition. Complementary tests of colliders, DM and GW from the minimal dark boson plasma  are shown in Sec.~\ref{sec:DM&GW}. Finally, conclusions are made in Sec.~\ref{sec:cons}. An analytic derivation of the dark phase transition is presented in Appendix~\ref{appen:analytic_PT} to efficiently  show the favored parameter space, and the  calculation of the  DM annihilation rate is relegated to   Appendix~\ref{appen:DM_rates}.

\section{A minimal Higgs-portal dark $U(1)_X$  with DM}\label{sec:framework}
\subsection{Theoretical framework}
The minimal dark gauge  $U(1)_X$ sector   contains only two  bosons, the dark scalar and the dark gauge boson. 
Here,  we consider a generic scalar potential without  classical scale invariance.  The tree-level scalar potential in vacuum is written as
 \begin{align}\label{eq:tree-level-potential}
 	V_{0}(S) =-\mu_S^{2}|S|^2+\lambda |S|^4\,,
 \end{align}
 where $\mu_S^{2}>0$ and  $S=(v_\phi+\phi+i\chi)/\sqrt{2}$ is the dark scalar singlet charged under $U(1)_X$ with $v_\phi$ the  vacuum expectation value (VEV).  $\phi$ is the   dark scalar and $\chi$ the Goldstone boson associated with the  gauge  $U(1)_X$ symmetry breaking.

 The dark boson plasma can exhibit $Z_2$-parity conservation, in which   the scalar and gauge boson transform as~\cite{Lebedev:2011iq,Farzan:2012hh}
 \begin{align}
 A'_\mu\to -A'_\mu\,, \quad S\to S^*\,.
 \end{align}
The  $Z_2$ symmetry forbids the kinetic mixing $B'_{\mu\nu}B^{\mu\nu}$, where $B'_{\mu\nu}=\partial_\mu A'_\nu -\partial_\nu A'_\mu$ and  $B_{\mu\nu}=\partial_\mu B_\nu -\partial_\nu B_\mu$ with $B_\mu$ the Abelian gauge fields in the SM sector. After the gauge $U(1)_X$ symmetry breaking, the $Z_2$ symmetry is retained, so that the dark gauge boson becomes a  stable DM candidate.
It should be mentioned that, however, both the dark scalar and the dark gauge boson can be   long-lived DM without the $Z_2$ parity, provided that they are lighter than $\sim 1$~MeV and the Higgs portal coupling as well as the kinetic mixing is sufficiently small. Nevertheless, when all the renormalizable portal couplings between the SM and the dark sectors are   sufficiently small, it becomes much more difficult to probe the dark in collider experiments. 

After the gauge $U(1)_X$ symmetry breaking, the gauge boson DM acquires mass via the dark vacuum $v_\phi$. Without fermions, there is freedom to normalize the gauge charge of the dark scalar such that, 
\begin{align}\label{eq:mA}
m_{A'}=g_X v_\phi\,,  
\end{align}
where $g_X$ denotes the dark gauge coupling. Since $v_\phi, g_X$ are relevant for the phase transition and GW, while the DM relic density and direct detection depend on $g_X$ and $m_{A'}$, we expect that   there exists some  correlation among the dark phase transition, the induced GW signals and the DM phenomenology, which we are about to investigate in this paper.

 \subsection{Higgs portal for  dark mediator detection}
In the presence of an unbroken $Z_2$ parity, the only renormalizable interaction between the minimal dark  and SM sectors comes from  the Higgs portal coupling, 
\begin{align}
\mathcal{L}\supset \lambda_p|S|^2 |H|^2, 
\end{align} 
where $H=(0,(v_{\rm EW}+h)/\sqrt{2})^T$ is the SM Higgs doublet with $v_{\rm EW}=246$~GeV the electroweak VEV.
While the portal interaction may   assist in   the electroweak phase transition,   an $\mathcal{O}(1)$  $\lambda_p$ is generically required~\cite{Vaskonen:2016yiu,Beniwal:2017eik,Hashino:2018zsi}. Since we are concerned with a dark phase transition,  we assume a small   portal coupling   such that it has a negligible  impact on the electroweak and dark phase transitions.  
 With   a  small $\lambda_p$, the   tree-level minimization condition from potential~\eqref{eq:tree-level-potential} approximately gives rise to 
\begin{align}\label{eq: tree-level_mass}
	 \mu_S^{2}\approx \lambda v^{2}_\phi\,, \quad m_{\phi}^{2}\approx 2\lambda v^{2}_\phi\,,
\end{align}
where $m_\phi$ is the physical mass of the dark scalar\footnote{In the presence of the Higgs portal coupling, the scalar  $\phi$ from the dark gauge $U(1)_X$ sector may be detected, though we will still call $\phi$ by the \textit{dark scalar} throughout.}. 
%
%
%
%
%
%

The Higgs portal coupling can play three important roles in the minimal dark sector, as we shall discuss in subsequent sections.  First of all,   the scalar-Higgs mixing after gauge symmetry breaking can induce couplings between the dark scalar and the SM particles.
For instance, the effective coupling between $\phi$ and the SM fermions is   given by $\theta m_f/v_{\rm EW}$,  
with $m_f$ the fermion mass and   $\theta$ the scalar-Higgs  mixing angle given by
\begin{align}\label{eq:lambda-theta}
	\theta\approx \frac{v_{\rm EW}v_\phi\lambda_p}{m_h^2-m_\phi^2}\,,
\end{align}
where  $\sin\theta\approx \theta$  is used and  $m_h=125.25$~GeV is the mass of the SM Higgs boson. 

The search of a light dark scalar with $m_\phi<5$~GeV  has been conducted and  proposed  in various experiments (see Refs.~\cite{Batell:2022dpx,Ferber:2023iso} for a recent review). For an electroweak-scale or higher dark phase transition,  $m_\phi$ can readily be above  $\mathcal{O}(10)$~GeV. The promising detection could then come from the Higgs invisible decay $h\to2\phi$, if $m_\phi<m_h/2$ and   $c\gamma \tau_\phi>1$~m, with  $\tau_\phi$  the lifetime of $\phi$  and   the decay width 
\begin{align}
\Gamma_{h\to2\phi}=\frac{\lambda_p^2 v_{\rm EW}^2 m_h^3}{32\pi(m_h^2-m_\phi^2)^2}\left(1+\frac{2m_\phi^2}{m_h^2}\right)^2\sqrt{1-\frac{4m_\phi^2}{m_h^2}}\,,
\end{align}
where  the dependence on $\theta$ is translated into that on the portal coupling via Eq.~\eqref{eq:lambda-theta}.
The current bound of the branching ratio  $\text{BR}(h\to 2\phi)=\Gamma(h\to 2\phi)/\Gamma_h$\footnote{$\Gamma_h=\Gamma_{h,\rm SM}+\Gamma_{h\to2\phi}$, with $\Gamma_{h,\rm SM}\approx 4.1$~MeV~\cite{Denner:2011mq}.} from  CMS  gives at $95\%$ C.L.~\cite{CMS:2018yfx}:
\begin{align}
	\text{BR}(h\to 2\phi)<19\%\,,
\end{align}
which, in the limit of $m_\phi\ll m_h$, is translated into an upper bound of the portal coupling: $\lambda_p<0.014$.  The detection from HL-LHC~\cite{deBlas:2019rxi}, CEPC~\cite{Tan:2020ufz} and ILC~\cite{Ishikawa:2019uda,Fujii:2020pxe,Potter:2022shg} will increase the sensitivity of the portal coupling by at least an order of magnitude. 

When $m_\phi>m_h/2$, the SM Higgs cannot decay into the dark scalar. In this case,  the  dark scalar can still be searched for via direct production, e.g.,  from proton-proton beams, followed by decay to SM particles.  The current bounds from 13~TeV CMS~\cite{CMS:2018amk,CMS:2021yci} and 13~TeV ATLAS~\cite{ATLAS:2018sbw,ATLAS:2020fry,ATLAS:2020tlo,ATLAS:2022xzm}  in the channels $pp\to \phi\to WW/ZZ/hh/\ell \ell$ have set  an upper limit on the mixing angle $\theta$ at the level of 0.1, while the future upcoming/proposed experiments will further increase the sensitivity of the mixing angle. 
 In particular, the future  14~TeV muon collider  with an integrated luminosity 14~$\text{ab}^{-1}$ can probe a mixing angle  at the level of 0.01  via the decay channel $\phi\to 2h$~\cite{Buttazzo:2018qqp}.

For $m_\phi>m_h/2$, on the other hand,  indirect signals of the heavy dark scalar from collider experiments can also be probed. A  simple and powerful way is to consider the signal/coupling strength modifier, defined via the cross section times the branching ratio in the narrow-width approximation~\cite{LHCHiggsCrossSectionWorkingGroup:2013rie},
\begin{align}
	\sigma(i\to h)\text{BR}(h\to f)\equiv \mu_i^f \times (\sigma_{i,\rm SM}\text{BR}_{h\to f,\rm SM})\equiv\frac{(\sigma_{i,\rm SM}\times \kappa_i^2)(\Gamma_{h\to f,\rm SM}\times \kappa_f^2)}{\Gamma_{h,\rm SM}\times \kappa_h^2}\,.
\end{align}
The signal  strength modifier $\mu_i^f$ with initial/final state $i/f$ characterizes the deviation from the SM prediction in a given channel $i\to h\to f$, while the coupling strength modifier $\kappa_i$ is used to describe the deviations of the SM Higgs boson couplings to the SM fermions/gauge bosons, with  $\kappa_h^2\equiv \sum_j \kappa_j^2 \Gamma_{h\to j,\rm SM}/\Gamma_{h,\rm SM}$.  In the dark $U(1)_X$ scenario, the scalar-Higgs mixing induces a universal coupling strength modifier: $\kappa_i^2=\cos\theta^2=\kappa_h^2$, and a universal signal strength modifier: $\mu_i^f=\cos\theta^2\equiv \mu$.
The current measurements of a universal $\mu$ from ATLAS and CMS set $\sin\theta\lesssim 0.3$ at $95\%$ C.L.~\cite{ATLAS:2022vkf,CMS:2022dwd}, while the future electron-positron colliders, such as ILC, CEPC and FCC-ee, can reach a $\kappa_Z$ precision of  the $ZZh$ coupling at  the level of $0.1\%$~\cite{deBlas:2019rxi,EuropeanStrategyforParticlePhysicsPreparatoryGroup:2019qin}. With the uncertainty at $2\sigma$ level, it implies that   a mixing angle down to the level of $0.06$ can be probed.

It is noteworthy that a universal coupling strength modifier is a characteristic feature for the dark scalar singlet, which can itself serve as a   distinguishable property  from variant scalar-extended scenarios, such as the two-Higgs-doublet models~\cite{Branco:2011iw,Kanemura:2014bqa}.

Before going to another  window associated with the direct detection of DM in Sec.~\ref{sec:direct_detection}, let us now turn to a fundamental setup for the dark evolution in the early universe, which is relevant for the dynamics of the dark phase transition and   the DM freeze-out.

\subsection{Higgs portal for   dark  thermalization}\label{sec:thermalization}
Besides opening the detection channels of the minimal dark sector at colliders, the Higgs portal coupling also plays an important role in   dark thermalization. In the early universe, the minimal dark sector could  share a common temperature with the SM thermal bath before the dark phase transition, if the portal coupling is sufficiently large.  It could also be the case that the dark sector has a different temperature from the SM one due to, e.g., an asymmetric reheating dynamics from the very beginning of the early universe after inflation, or an early decoupling before the dark phase transition. However, when the dark sector is colder than the SM plasma, with a temperature ratio $\xi\equiv T_{\rm D}/T_{\rm SM}<1$,  the vacuum energy release during  the first-order   phase transition would be suppressed by $\xi^4$~\cite{Breitbach:2018ddu}, and in particular, the sound-wave dominated  GW peak amplitude would be suppressed by $\xi^8$~\cite{Li:2023bxy}, making the GW signal harder to detect. Therefore, a large enough portal coupling  is favored to  ensure a hot dark plasma that can  generate strong GW signals.     

The lower limit of $\lambda_{p}$ for a common temperature ($\xi=1$, or one-temperature treatment) between the dark and the SM sectors can be derived as follows. Before the electroweak and dark gauge symmetry breaking, the dominant thermalization process
comes from the quartic scalar-Higgs scattering: $2H\to2S$.  The collision rate can be written as~\cite{Gondolo:1990dk}:  
\begin{align}
	\gamma_{2H\to2S}  =\frac{T}{32\pi^4}\int_{\hat{s}_{\rm min}}^{\infty}d\hat{s}\sqrt{\hat{s}}(\hat{s}-4m_{H}^{2})K_{1}(\sqrt{\hat{s}}/T)\sigma_{2H\to2S}\, ,
\end{align}
where $\hat{s}_{\rm min}=\max [4m_{H}^{2},4m_S^2]$ with $m_H\approx 0.4 T$~\cite{Cline:1993bd}  the SM Higgs thermal  mass and $m_S^2(T)\approx g_X^2 T^2/4$ the thermal mass of the dark scalar\footnote{This can be derived by $\partial^2 V_{{\rm eff}}(\varphi,T)/\partial \varphi^2$ with $V_{{\rm eff}}(\varphi,T)$ given in~\eqref{eq:potential-1} or~\eqref{eq:effpotential}.}. $K_1$ is the modified Bessel function, and the   cross section is given by
\begin{align}
	\sigma_{2H\to2S} & =\frac{\lambda_{p}^{2}\beta_S}{16\pi \beta_H \hat s}\,,
\end{align}
where $\beta_{S,H}\equiv \sqrt{1-4m_{S,H}^2(T)/\hat{s}}$. The thermalization condition for the one-temperature treatment of the dark phase transition requires that the thermally averaged rate $\langle\sigma v \rangle n\equiv\gamma_{2H\to2S}/n_{\rm eq}$, with $n_{\rm eq}\approx 0.12 T^3$   the thermal scalar number density, should  be  larger than    the Hubble rate,  
\begin{align}\label{eq:Hubble}
	\mathcal{H}=\sqrt{\frac{4\pi^{3}}{45}g_{\rho}(T)}\frac{T^{2}}{M_{\rm Pl}}\,,
\end{align}
before the phase transition completes. Here, $g_{\rho}(T)$ is the relativistic degrees of freedom for energy density and $M_{\rm Pl}=1.22\times 10^{19}$~GeV the Planck mass. A lower bound of the portal coupling can then  be derived by  $\langle\sigma v \rangle n>\mathcal{H}$.  Neglecting the $\beta_{S}$ dependence, we obtain a simple lower bound of $\lambda_{p}$ in terms of the temperature,
\begin{align}
	\lambda_{p}\gtrsim 9.5 \times10^{-8}\left(\frac{T}{{\rm GeV}}\right)^{1/2},\label{eq:lower_portal}
\end{align}
where $g_\rho(T)=110.75$ is used as a reference. From Eq.~\eqref{eq:lower_portal}, we can see that for an electroweak-scale dark phase transition, $\lambda_{p}>\mathcal{O}(10^{-6})$ is required. 

After the dark phase transition and the gauge  symmetry breaking, the dark scalar can either be in thermal equilibrium or go out of equilibrium. Requiring the SM and the DM to be still in thermal equilibrium prior to the DM
freeze-out is the simplest and most minimal setup to derive the DM dynamics, as will be considered here.  
The   DM freeze-out in the one-temperature treatment requires that before the freeze-out temperature at $T_{\rm fo}\simeq m_{A'}/25$, the dark scalar should keep $A'_\mu$ in thermal equilibrium via scalar-SM interactions. Depending on $T_{\rm fo}$ and $m_\phi$, the dominant process keeping $\phi$ in thermal contact with the SM plasma can  come either from two-body scalar-fermion or from two-body scalar-boson process, such as $\phi  \leftrightharpoons b\bar b$ or $\phi  \leftrightharpoons 2h$. To follow the one-temperature freeze-out, the thermally averaged  decay  rate   should be larger than the Hubble parameter  before $T_{\rm fo}$.

For $T_{\rm fo}\sim \mathcal{O}(1-10)$~GeV and $m_h>m_\phi>m_h/2$,   we  evaluate the thermal condition via  $\phi  \leftrightharpoons b\bar b$, with the vacuum decay width  
\begin{align}
	\Gamma_{\phi\to b \bar b}=\frac{3\theta^2m_b^2}{8\pi v_{\rm EW}^2}m_\phi \left(1-\frac{4m_b^2}{m_\phi^2}\right)^{3/2},
\end{align}
where $m_b$ is the mass of the bottom quark.
The thermally averaged decay rate  reads
 \begin{align}
\langle \Gamma_{\phi\to b\bar b}\rangle &\equiv \frac{1}{n_{\phi,\rm eq}}\int   d\Pi_{\phi}d\Pi_{b} d\Pi_{\bar b} f_{\phi}  |\mathcal{M}|_{\phi\to b\bar b}^2 (2\pi)^4 \delta^4\left(p_\phi-p_b-p_{\bar b}\right)
\\[0.2cm]
&\approx\Gamma_{\phi\to b\bar b}\frac{K_1(m_\phi/T)}{K_2(m_\phi/T)}\,,
 \end{align}
where $d\Pi_{i}\equiv d^3p_i/[(2\pi)^32 E_i]$ is the phase-space factor and $n_{\phi,\rm eq}$ denotes the thermal number density of the dark scalar.  The second line in the above equation  is obtained by taking the Boltzmann approximation $f_\phi=e^{-E_\phi/T}$.\footnote{When the complementary tests are concerned in Sec.~\ref{sec:DM&GW}, this approximation suffices to derive the bounds of the Higgs portal coupling.}  

For  $T_{\rm fo}\gg  10$~GeV and $m_\phi>2m_h$,
 we  evaluate the thermal condition via  $\phi \leftrightharpoons 2h$, with the vacuum decay width
\begin{align}
	\Gamma_{\phi\to 2h}=\frac{\theta^2m_\phi^3}{32\pi v_{\rm EW}^2} \left(1-\frac{4m_h^2}{m_\phi^2}\right)^{1/2}\left(1+\frac{2m_h^2}{m_\phi^2}\right)^2.
\end{align}
The thermally averaged decay   rate reads
\begin{align}
	\langle \Gamma_{\phi\to 2h}\rangle &\equiv \frac{1}{n_{\phi,\rm eq}}\int   d\Pi_{\phi}d\Pi_{h} d\Pi_{h} f_{\phi}  |\mathcal{M}|_{\phi\to 2h}^2 (2\pi)^4 \delta^4\left(p_\phi-p_h-p_{h}\right)
	\\[0.2cm]
	&\approx\Gamma_{\phi\to 2h}\frac{K_1(m_\phi/T)}{K_2(m_\phi/T)}\,.
\end{align}
The one-temperature condition for the DM freeze-out then requires
\begin{align}
\frac{\langle \Gamma_{\phi\to b\bar b}\rangle}{\mathcal{H}}\bigg |_{T=T_{\rm fo}}>1\,,\quad  \text{or}\quad \frac{\langle \Gamma_{\phi\to 2h}\rangle}{\mathcal{H}}\bigg |_{T=T_{\rm fo}}>1\,.
\end{align}


Thus far, we have seen that   a large enough portal coupling is favored  for  detection and thermalization of the dark sector. 
Besides, when  the gauge boson is a stable DM candidate, the Higgs portal also induces interactions between the SM particles and the gauge boson, and in particular, the interaction between nucleon and DM.  We will show in Sec.~\ref{sec:direct_detection} that the DM direct detection   via nucleon-DM scattering also sets severe bounds on the Higgs portal coupling, while the future sensitivity from, e.g.,  the  LZ experiment will be able to test such a minimal dark sector. 

\section{Dark gauge boson as WIMP-like DM}\label{sec:WIMPDM}
\subsection{Relic density}
\begin{figure}[t]
	\centering
	\includegraphics[scale=0.9]{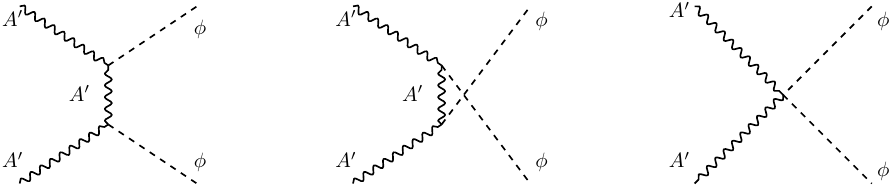}
	\caption{The dominant annihilation processes for the gauge boson DM freeze-out.
		\label{fig:2X2phi}
	}
\end{figure}

If the dark scalar is thermalized via the Higgs portal coupling, the strong interaction in the dark sector  can further keep the dark gauge boson in thermal equilibrium. Afterwards, the gauge boson DM undergoes freeze-out via annihilation to the dark scalar. 
The freeze-out process to determine the DM relic density  is   via  $2A'_\mu \to 2\phi$, with  the Feynman diagrams   shown in Fig.~\ref{fig:2X2phi}. It is noteworthy that the vector DM $A'_\mu$ can also have a portal interaction with the SM Higgs via $h^2A^\prime_\mu A^{\prime \mu}$, which was considered in Refs.~\cite{Kanemura:2010sh,Lebedev:2011iq,Baek:2012se} as the dominant portal  for a vector DM and could be induced by integrating out a heavy dark scalar at  tree level via the Higgs portal. Since we are interested in a small Higgs portal coupling, the dominant  processes to determine the DM freeze-out reduce to the dark interaction  shown in Fig.~\ref{fig:2X2phi}.

The    Boltzmann equation for the dark gauge boson freeze-out is given by
\begin{alignat}{1}
	\frac{dY_{A'}}{dx} & =-\frac{s\langle\sigma v\rangle_{A'}}{\mathcal{H}x}\left(Y_{A'}^{2}-\frac{Y_{\phi}^{2}}{Y_{\phi,{\rm eq}}^{2}}Y_{A',{\rm eq}}^{2}\right),\label{eq:YV}
\end{alignat}
where the  yield  is defined as $Y_{A',\phi}\equiv n_{A',\phi}/s$ and $x\equiv m_{A'}/T$.  The   entropy density is given by
\begin{align}\label{eq:s_and_H}
	s  =\frac{2\pi^{2}}{45}g_{s}(T)T^3\, ,
\end{align} 
with  $g_{s}(T)$ the relativistic degrees of freedom for entropy density in the plasma, and   the Hubble parameter $\mathcal{H}$ is given by Eq.~\eqref{eq:Hubble}.
 The 
thermally averaged cross section of DM annihilation can be defined as
\begin{align}
	\langle\sigma v\rangle_{A'} 
	\equiv 2(a+3b/x)\, ,\label{eq:sigmav-2V2phi}
\end{align}
where   we add a factor of 2    to take into account the pair DM annihilation, with the expressions of $a$ and $b$   collected  in   Appendix~\ref{appen:DM_rates}. 

Under the one-temperature regime, the scalar is still in thermal equilibrium when the gauge boson starts
to freeze out. Therefore we  take $Y_{\phi}\approx Y_{\phi,{\rm eq}}$
in Eq.~\eqref{eq:YV}. Following the standard semi-analytic approach~\cite{Kolb:1990vq},
we can parameterize the departure
from equilibrium by  $\Delta \equiv Y_{A'}-Y_{A',{\rm eq}}$. The Boltzmann equation for $\Delta$ is then given by 
\begin{align}
	\frac{d\Delta }{dx} & =-\frac{dY_{A',{\rm eq}}}{dx}-\frac{s\langle\sigma v\rangle_{A'}}{\mathcal{H}x}\Delta [\Delta +2Y_{A',{\rm eq}}]\, .\label{eq:dDeltadx}
\end{align}
Before the gauge boson departs from chemical equilibrium, the change
of $\Delta $ is small, so that we can write
\begin{align}\label{eq:Delta}
	\Delta  = -\frac{dY_{A',{\rm eq}}}{dx}\frac{\mathcal{H}x}{s\langle\sigma v\rangle_{A'}}\frac{1}{\Delta+2Y_{A',{\rm eq}}} \, .
\end{align}
During the freeze-out, $Y_{A'}$ departs significantly from $Y_{A',\rm eq}$ so that $\Delta$ is comparable to $Y_{A',{\rm eq}}$. Defining
$\Delta \equiv c Y_{A',{\rm eq}}$ at freeze-out,  with an order-one
constant $c$ determined by   numerical matching, we can derive the freeze-out temperature from Eq.~\eqref{eq:Delta} as
\begin{align}\label{eq:YVf-2}
	x_{f}=\ln\left(\sqrt{\frac{45}{2}}\frac{c_{}(c_{}+2)}{4\pi^{3}}\frac{g_{s}g_{A'}}{g_{\rho}^{3/2}}M_{\rm Pl}m_{A'}\frac{\langle\sigma v\rangle_{A'}}{\sqrt{x_{f}}[1-3/(2x_{f})]}\right),
\end{align}
where $g_{A'}=3$ is the internal degrees of freedom for the massive gauge boson and we have used the thermal yield of the nonrelativistic $A'_\mu$,
\begin{align}\label{eq:YVeq}
	Y_{A',{\rm eq}}=\frac{45g_{A'}x^{3/2}e^{-x}}{4\sqrt{2}\pi^{7/2}g_s(T)}\, .
\end{align}
Towards the end of freeze-out, $\Delta$ turns to be constant while $Y_{A', \rm eq}$ becomes negligible due to Boltzmann suppression. During this stage, we can obtain a simple  equation of $\Delta$ from Eq.~\eqref{eq:dDeltadx} as
\begin{align}
 d\Delta^{-1}=\frac{s\langle\sigma v\rangle_{A'}}{\mathcal{H}x}dx\,.
\end{align}
The final abundance of $A'_\mu$ is given by   integrating $x$ in the above equation from $x=x_f$ to $x=\infty$, leading to  
\begin{align}\label{eq:YA_fin}
	Y_{A'}(x=\infty)\approx \Delta (x=\infty)=\left(\int_{x_{f}}^{\infty}\frac{s\langle\sigma v\rangle_{A'}}{\mathcal{H}x} dx +[c Y_{A',{\rm eq}}]^{-1}\Big|_{x_{f}}\right)^{-1}\,.
\end{align}
The relic density is given by
\begin{align}
	\Omega_{\rm DM}h^{2} & =\frac{s_{0}m_{A'}}{\rho_{c}/h^{2}}Y_{A'}(x=\infty)\, ,
\end{align}
with the current entropy density $s_{0}=2891.2~{\rm cm^{-3}}$ and the critical density today $\rho_{c}=1.05\times10^{-5}\cdot h^{2}\cdot {\rm GeV \cdot {\rm cm^{-3}}}$~\cite{Workman:2022ynf}. 

Since a strong first-order phase transition favors $m_{A'}> m_\phi$, the DM relic density depends essentially on $m_{A'}$ and $g_X$ only. Matching the observed value $\Omega h^2\approx 0.12$~\cite{Planck:2018vyg} then allows us to remove one of these free parameters.  
To this end, we approximate the freeze-out temperature   as 
\begin{align}\label{eq:xf_approx}
x_f\approx 42.83+\ln\left(\frac{g_X^4}{\sqrt{g_{\rm eff}}}\frac{\rm GeV}{m_{A'}}\right),
\end{align}
where we have taken $c\approx 0.4$~\cite{Kolb:1990vq},  $x_f\approx 25$ and $g_{\rm eff}\equiv g_s\approx g_\rho$ in the logarithmic function of Eq.~\eqref{eq:YVf-2}, though $x_f$ can be numerically solved by iteration.  It can be seen that for an electroweak $A'_\mu$ and a weak coupling $g_X$, the freeze-out temperature is expected to be around $x_f\approx 25$. 

The relic density can be approximated as
\begin{align}\label{eq:omega_approx}
	\left(\frac{\Omega_{\rm DM}h^2}{0.12}\right)\approx 0.33 \left(\frac{0.1}{g_X}\right)^4  \left(\frac{m_{A'}}{100~\rm GeV}\right)^2,
\end{align}
where we have taken  a benchmark value $g_{\rm eff}=80$,   corresponding  to a freeze-out temperature around $\mathcal{O}(10-100)$~GeV and hence $m_{A'}\simeq \mathcal{O}(100-1000)$~GeV. We will use Eq.~\eqref{eq:omega_approx}, or equivalently 
\begin{align}\label{eq:mAofg}
m_{A'}\approx 1.75\times 10^4 g_X^2~\text{GeV}\,, \quad v_\phi\approx   1.75\times 10^4 g_X~\text{GeV}\,,
\end{align}
as a numerical approximation to remove the dependence of the dark phase transition on $m_{A'}$ and  $v_\phi$.

 As will be shown in this paper,  $0.1<g_X<1$ and $10^{-4}<\lambda\ll 1$ are favored for a strong first-order dark phase transition and observable GW signals. Under this circumstance, the DM mass is predicted to be around  100~GeV$-$10~TeV, which is below the perturbative unitarity bound at $\mathcal{O}(100)$~TeV~\cite{Griest:1989wd}, while the high-energy elastic scattering matrices for the longitudinal  gauge boson (or equivalently $\chi$) and the dark scalar $\phi$ can also readily satisfy the unitarity constraint due to a small quartic coupling $\lambda$~\cite{Lee:1977eg}. 

\subsection{Higgs portal for DM direct detection}\label{sec:direct_detection}

\begin{figure}[t]
	\centering
	\includegraphics[scale=1]{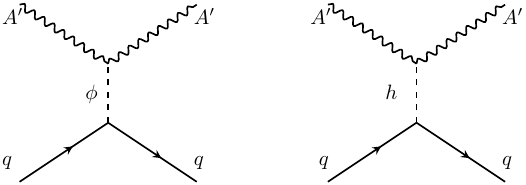}
	\caption{The DM-nucleon scattering mediated by the dark scalar $\phi$ and the SM Higgs boson $h$, where $q$ denotes the quark component in the nucleon.
		\label{fig:2A2q}
	}
\end{figure}
 The dark gauge boson DM 
can interact with the SM fermions via the scalar-Higgs  mixing induced
coupling.  The  direct detection via  scattering between DM and nucleons has been conducted in   LUX~\cite{LUX:2016ggv}, XENON1T~\cite{XENON:2017vdw}, PandaX-II~\cite{PandaX-II:2017hlx}, PandaX-4T~\cite{PandaX-4T:2021bab} and  LZ~\cite{LZ:2022ufs} experiments.  The null observations from these experiments have set severe bounds in the spin-independent cross section in terms of the DM mass.  Future    sensitivity from LZ~\cite{LZ:2015kxe,Mount:2017qzi} and DARWIN~\cite{Baudis:2012bc,Schumann:2015cpa} will further  improve the cross section
limit by at least  two orders of magnitude.   

The scattering between the DM $A'_\mu$ and the nucleons goes through the $t$ channel  mediated by the two physical scalar states, i.e., $\phi$ and $h$, as shown in Fig.~\ref{fig:2A2q}. 
The spin-independent cross section is straightforward to obtain, giving
\begin{align}\label{eq:signma_AN}
	\sigma_{A'N}=\frac{f^2 g_X^4  v_\phi^2\theta^2}{\pi v_{\rm EW}^2}\frac{m_N^4 (m_h^2-m_\phi^2)^2 }{m_h^4 m_\phi^4 (m_N+m_{A'})^2}
= \frac{f^2 \lambda_{p}^2m_{A'}^4m_N^4}{\pi m_h^4 m_\phi^4(m_N+m_{A'})^2}\,,
\end{align}
where $m_N$ is the nucleon mass\footnote{Usually the  measured DM-nuclear cross section  has been reduced by defining an effective per-nucleon cross section so as to facilitate comparison between different experiments with different nuclei (such as xenon in the  LZ experiment)~\cite{Lewin:1995rx}. Therefore, we will take the nucleon mass  $m_N\approx 0.931$~GeV to evaluate the effective DM-nucleon cross section in Eq.~\eqref{eq:signma_AN} without the quadratic scaling of the atomic mass number,  and compare with the excluded    cross section.} and we use  $f\approx 0.3$~\cite{Hill:2014yxa,Hoferichter:2017olk} as the effective Higgs-proton (neutron) coupling from the nucleon matrix elements $f\equiv \sum_q \langle N|m_q q\bar q/m_N|N\rangle$.
Note that, we have neglected the recoil energy dependence in the dark scalar propagator. This approximation will be justified in Sec.~\ref{sec:DM&GW} where a strong first-order phase transition typically  predicts  $m_\phi>\mu_{A'N} E_R$, with $\mu_{A'N}\equiv m_{A'}m_N/(m_{A'}+m_N)$  the  reduced DM-nucleon mass and $E_R$ the     recoil energy.

For the one-temperature freeze-out, the dark gauge boson acts as a DM particle between  the standard WIMP and the secluded WIMP paradigms~\cite{Pospelov:2007mp}, since the annihilation  product  is  within the dark sector rather than the SM particles and the coupling fixing the relic density is largely disentangled from that controlling the DM-nucleon cross section.   Under this circumstance,   the tension between the null observations and the standard WIMP prediction can readily be alleviated via a small mixing angle $\theta$.  However, since a sufficiently large portal coupling is favored for detection and strong GW signals, it is interesting to investigate the  favored ranges of  $\lambda_p$ which  can  explain the current null observations and at the same time  predict a DM-nucleon cross section   detectable by   future experiments   above the notorious neutrino floor~\cite{Billard:2013qya,Ruppin:2014bra}.  This is one of the main purposes in this paper, as we shall discuss in Sec.~\ref{sec:DM&GW}.

\subsection{Higgs portal for DM indirect detection}\label{sec:DM-indirect}
As mentioned above, the dominant DM annihilation channel is the dark scalar product, while the cross sections of   annihilation to SM particles are expected to be suppressed by the portal coupling $\lambda_{p}$ and the mass ratio $m_{{\rm SM}}/m_{A'}$. In this case, a significant indirect signal of DM annihilation at present day could still be  feasible  if the annihilation cross sections are boosted by the Sommerfeld enhancement~\cite{Sommerfeld:1931qaf} for low DM velocities ($v\simeq\mathcal{O}(10-1000)\text{km/{\rm s}}$) in the regions of typical dwarf galaxies or galaxy clusters~\cite{Tulin:2017ara}. 

The Sommerfeld effect is  a non-relativistic quantum effect, arising from a force carrier between the incoming particles. The force carrier, or the mediator between the two nonrelativistic particles,  can distort the incoming wave functions away from the plane-wave approximation, potentially yielding enhancement or suppression to annihilation cross section. In the minimal $U(1)_X$ scenario, the dark scalar $\phi$ plays the role of mediator between two gauge bosons $A_\mu'$.
For the Sommerfeld effect to be significant, however, the mass hierarchy between  $\phi$ and $A_\mu'$ should typically  satisfy the following condition~\cite{Hisano:2004ds,Arkani-Hamed:2008hhe,Lattanzi:2008qa,Hryczuk:2010zi}:
\begin{align}\label{eq:mass-ratio}
m_\phi\lesssim \alpha_X m_{A'}\,,
\end{align} 
where $\alpha_X\equiv g_X^2/(4\pi)$. The above condition reduces to 
\begin{align}\label{eq:coupling-ratio}
\lambda\lesssim \frac{g_{X}^{6}}{32\pi^{2}}\,.
\end{align}
So one can expect that the Sommerfeld effect only arises when the self-interacting coupling of the dark scalar is much smaller than the gauge coupling. 

To estimate the   DM annihilation cross section to SM particles,  let us consider the dominant channel $2A'_\mu\to t\bar t$. The process is mediated by the Higgs boson and the dark scalar in the $s$ channel,  the cross section of which reads
\begin{align}\label{eq:A-SM}
	(\sigma v)_{A'\to t}\approx \frac{\lambda_p^2 m_t^2}{512\pi m_{A'}^4}\propto \left(\frac{\lambda_{p}}{g_X^2}\right)^2 \left(\frac{m_{t}}{m_{A'}}\right)^2(\sigma v)_{A'\to \phi}\,.
\end{align}
The above result corresponds to the $s$-wave approximation in the limit $m_{A'}>m_\phi, m_t$,  with $m_t$ the top-quark mass. Besides, we have used Eq.~\eqref{eq:lambda-theta} and Eq.~\eqref{eq:sigmav} for $(\sigma v)_{A'\to \phi}$.  We can see that, comparing the $2A'_\mu\to 2\phi$, 	$(\sigma v)_{A'\to t}$ would be much suppressed for  $m_{A'}\gg m_{t}$ and $g_X^2\gg \lambda_p$.  

Therefore, without a significant Sommerfeld enhancement, the indirect signals of DM via annihilation to SM particles or via annihilation to dark scalar followed by   decay to SM particles can hardly be detected.  We will find out  in  Sec.~\ref{sec:DM&GW} that   Eq.~\eqref{eq:mass-ratio} or Eq.~\eqref{eq:coupling-ratio} is in general not satisfied when observational GW signals are concerned.

\section{Dark phase transition from a minimal hot boson plasma}\label{sec:PT}
The dark scalar potential~\eqref{eq:tree-level-potential} at finite temperatures in the early universe could exhibit a global minimum at $\langle S\rangle=0$ rather than   $\langle S\rangle=v_\phi$ such that the gauge $U(1)_X$ symmetry is maintained above some critical temperature $T_c$. At $T_c$, which is defined by the epoch when there are degenerate vacua  such that $V(\langle S\rangle_1)=V(\langle S\rangle_2)$, another  minimum of the potential at $\langle S\rangle\neq 0$ arises. The two degenerate vacua are separated by an energy barrier, which is a necessary condition for a strong first-order phase transition. At some temperatures just below $T_c$, the  quantum/thermal tunneling from the false to true vacuum occurs, and a first-order phase transition begins,  followed by  spontaneous gauge  symmetry breaking.  A common practice for the dynamics of the phase transition is the effective potential analysis, as we shall describe in this section.
\subsection{Effective potential}
The effective potential in the dark boson plasma  can be generically
written as
\begin{alignat}{1}
V_{{\rm eff}}(\varphi,T) & =V_{0}+V_{{\rm CW}}+V_{\rm CT}+V_{T}+V_{\rm daisy}\,,\label{eq:potential-1}
\end{alignat}
where the tree-level potential for the background field $\varphi$ at zero temperature is given by
\begin{alignat}{1}
V_{0}(\varphi) & =-\frac{1}{2}\mu_S^{2}\varphi^{2}+\frac{\lambda}{4}\varphi^{4}\,,\label{eq:tree-level-background-potential}
\end{alignat}
and  the tree-level $\varphi$-dependent masses of the quantum fields read
\begin{align} 
		m_{\phi}^{2}(\varphi)=-\mu_S^{2}+3\lambda\varphi^{2}\,,\quad m_{\chi}^{2}(\varphi)=-\mu_S^{2}+\lambda\varphi^{2}\,, & \quad m_{A'}^{2}(\varphi)=g_X^{2}\varphi^{2}\,.\label{eq:field-dependent-mass}
\end{align}

The zero-temperature Coleman-Weinberg potential   $V_{{\rm CW}}(\varphi)$
is induced by the quantum scalar fields $\phi,\chi$ and the gauge
vector field $A'_\mu$. In the  Landau
gauge with the  $\overline{{\rm MS}}$ renormalization scheme, $V_{{\rm CW}}(\varphi)$ reads~\cite{Coleman:1973jx}
\begin{align}\label{eq:VCW}
V_{{\rm CW}}(\varphi)& =\frac{1}{64\pi^{2}}\sum_{i=\phi,\chi,A'} n_i m_{i}^{4}(\varphi)\left[\ln\left(\frac{m_{i}^{2}(\varphi)}{\mu^{2}}\right)-C_i\right],
\end{align}
where  $\mu$ is the renormalization scale\footnote{We fix the scale by $\mu=v_\phi$ throughout. Since $\lambda$ and $g_X$ are treated as free parameters in this paper, we will not consider the renormalization dependence of $V_{\rm eff}(\varphi,T)$ in numerical analyses. }, $n_\phi=n_\chi=1, n_{A'}=3$ accounts for the internal degrees of freedom, and  $C_{\phi,\chi}=3/2,C_X=5/6$ are the   renormalization-dependent constants. 

There are notorious IR issues associated with the massless Goldstone boson $\chi$ in the   Coleman-Weinberg loop function. It can be clearly seen that when the tree-level mass of $m_\chi$ from Eq.~\eqref{eq:field-dependent-mass} in vacuum $\varphi=v_\phi$ is taken, there would be a divergence in the logarithmic function. It can also be seen that when  the vacuum mass of $\phi$ is determined  by adding loop corrections,  i.e., $\partial^2(V_0+V_{\rm CW})/\partial \varphi^2|_{\varphi=v_\phi}=m_\phi^2$, we will encounter the logarithmic divergence due to the second derivative of $V_{\rm CW}$ at $\varphi=v_\phi$. While there are several ways to evade a  massless Goldstone boson~\cite{Cline:1996mga,Elias-Miro:2014pca,Martin:2014bca}, we will simply drop the contribution from $\chi$~\cite{Espinosa:2011ax} in the following analysis. This approximation is reasonable for a minimal dark boson plasma, where the dominant contribution to induce a strong first-order phase transition comes from the gauge boson, with $\lambda\ll g_X^2$.

The counterterm $V_{\rm CT}\equiv -\delta\mu_S^2 \varphi^2/2+\delta\lambda^4\varphi^4/4$ is chosen to maintain  the tree-level vacuum and mass conditions of $\phi$, such that 
\begin{align}
\frac{	\partial (V_{\rm CW}+V_{\rm CT})}{\partial \varphi}\Big|_{\varphi=v_\phi}=0\,, \quad \frac{	\partial^2 (V_{\rm CW}+V_{\rm CT})}{\partial \varphi^2}\Big|_{\varphi=v_\phi}=0\,,
\end{align}
leading to 
\begin{align}
	V_{\rm CT}&=\frac{3v_\phi^2}{64\pi^2}\left[2g_X^4+2\lambda^2 \ln\left(\frac{m_\phi^2}{\mu^2}\right)+7\lambda^2\right] \varphi^2
	\nonumber \\[0.2cm]
	&-\frac{3}{128\pi^2}\left[3g_X^4+2g_X^4 \ln\left(\frac{m_{A'}^2}{\mu^2}\right)+9\lambda^2+6\lambda^2\ln\left(\frac{m_\phi^2}{\mu^2}\right)\right]\varphi^4,
\end{align} 
where  the masses in the logarithmic function are given by Eqs.~\eqref{eq:mA} and~\eqref{eq: tree-level_mass}.

The finite-temperature potential from $\phi,
\chi$ and $A'_\mu$ is given
by
\begin{alignat}{1}
V_{T} & =\frac{T^{4}}{2\pi^{2}}\sum_{i=\phi,\chi,A'} n_i J_{B}(x_i),\label{eq:VT-1}
\end{alignat}
where $J_B$ reads
\begin{align}\label{eq:JBint}
	J_B(x_i)=\int_0^\infty y^2 dy \ln\left(1-e^{-\sqrt{y^2+x_i^2}}\right),
\end{align}
with $x_i\equiv m_i(\varphi)/T$. Note that there is no fermion integral $J_F$ in $V_T$ as we consider a minimal dark boson plasma. For the numerical analysis, we keep  $V_T$ without expanding in the high-temperature limit. An analytic analysis with the high-temperature expansion will be presented in Appendix~\ref{appen:analytic_PT}.

Besides the finite-temperature corrections presented in Eq.~\eqref{eq:VT-1}, there are significant   corrections from daisy rings.  
Following  the Arnold-Espinosa resummation~\cite{Arnold:1992rz},  we can write the daisy potential from the gauge boson as
\begin{align}
	V_{\rm daisy}=-\frac{g_X^3T}{12\pi}\left[\left(\varphi^2+\frac{T^2}{3}\right)^{3/2}-\varphi^3\right],
\end{align}
which corresponds to the mass replacement: 
\begin{align}\label{eq:AEresum}
	m_{A',T(L)}^{2}(\varphi, T=0) \to m_{A',T(L)}^{2}(\varphi, T=0)+c_{T(L)}g_X^{2}T^{2}\,,
\end{align}
with $c_{T}=0,c_{L}=1/3$ for the transverse and longitudinal components of $A'_\mu$~\cite{Chiang:2017zbz,Hashino:2018zsi}. The thermal mass corrections to scalars can also be included similarly.  Nevertheless, we will neglect the scalar contributions to the dark phase transition, which is a good approximation as long as the gauge coupling $g_X$ is much larger than the scalar coupling $\lambda$. 

\subsection{Dynamics of dark phase transition}
During the phase transition, the true-vacuum bubbles start  to nucleate when the thermal tunneling probability per Hubble time and per Hubble volume  is at order unity. The nucleation temperature $T_n$ can be  defined by~\cite{Linde:1980tt,Linde:1981zj}:
\begin{align}\label{eq:Tn-def-1}
	\Gamma(T_n)\simeq T_n^4 e^{-S_3/T_n}\simeq \mathcal{H}^4(T_n)\,,
\end{align}
where the Hubble parameter is given by Eq.~\eqref{eq:Hubble}.
$S_3$ is the three-dimensional Euclidean action, 
\begin{align}\label{eq:EclideanS3}
	S_{3}=4\pi\int_{0}^{\infty}r^{2}dr\left[\frac{1}{2}\left(\frac{{d\varphi}}{dr}\right)^{2}+V_{\rm eff}(\varphi,T)\right],
\end{align} 
where the effective potential is given by Eq.~\eqref{eq:potential-1} and the background field $\varphi(r)$ is the $O(3)$-symmetric bounce solution to the equation of motion:
\begin{align}\label{eq:bounce-equation-1}
	\frac{{d^{2}\varphi}}{dr^{2}}+\frac{2}{r}\frac{d\varphi}{dr}=\frac{dV_{{\rm eff}}}{d\varphi}\,,
\end{align}
satisfying the boundary conditions
\begin{align}\label{eq:boundary-condition}
	\lim\limits_{r\to \infty}\varphi(r)=0\, ,\quad \frac{d\varphi(r)}{dr} \Big|_{r=0}=0\, .
\end{align}

The nucleation temperature from Eq.~\eqref{eq:Tn-def-1} can be rewritten as 
\begin{align}\label{eq:S3overT-1}
	\frac{	S_3(T_n)}{T_n}=146-2\ln\left(\frac{g_{\rho}(T_n)}{100}\right)-4\ln\left(\frac{T_n}{100~\rm GeV}\right).
\end{align}
To compute the nucleation temperature,  one should solve the bounce equation~\eqref{eq:bounce-equation-1}, which in general can be done by numerical approaches. To this end, we use  the Python package \texttt{CosmoTransitions}~\cite{Wainwright:2011kj} to calculate the Euclidean action $S_3$ and then determine the nucleation temperature via Eq.~\eqref{eq:S3overT-1}\footnote{Numerically, we fix  $g_\rho (T_n)=110.75$ and  $T_n=100$~GeV  in the logarithmic functions of Eq.~\eqref{eq:S3overT-1}  to get a fast evaluation of $T_n$  in \texttt{CosmoTransitions}. The resulting $T_n$   different from  $T_n=100$~GeV by   a factor of few is bearable due to the moderate logarithmic dependence. }. 

When associated with the GW profiles, the strength of the  phase transition is usually characterized by the  $\alpha$ parameter, which is defined by  the  difference of the  trace anomaly from the total energy-momentum tensor over the energy density in the symmetric phase~\cite{Hindmarsh:2015qta,Caprini:2019egz}:
\begin{align}\label{eq:alpha}
	\alpha= \frac{1}{\rho_R}\left( \Delta V_{\rm eff}-\frac{T}{4} \frac{\partial \Delta V_{\rm eff}}{\partial T}\right),
\end{align}
where the radiation energy  density $\rho_R$ in the plasma is given  by 
\begin{align}
\rho_R=\frac{\pi^2 g_\rho(T)}{30} T^4\,,
\end{align}
and $\Delta V_{\rm eff}$ is the potential difference between the false and true vacua: $V_{\rm eff}(\varphi_{\rm false},T)-V_{\rm eff}(\varphi_{\rm true},T)>0$.


Another important parameter describing the GW profiles is the so-called  $\beta$-parameter, which  is defined by the time derivative of the exponent in the thermal tunneling probability per unit time and per unit volume, i.e.,
\begin{align}
	\beta=-\frac{d}{dt}\left(\frac{S_3}{T}\right)=\mathcal{H}T\frac{d}{dT}\left(\frac{S_3}{T}\right)\, ,
\end{align}
where the relation $dT/dt=-\mathcal{H} T$ at the  radiation-dominated epoch is used in the second equation. Then the ratio   $\beta/\mathcal{H}$ at $T_n$ can be expressed as
\begin{align}\label{eq:beta/H}
	\frac{\beta}{\mathcal{H}}=T\frac{d}{dT}\left(\frac{S_3}{T}\right)\Big|_{T_n}\,.
\end{align}  
Note that    the  
 GW production is usually evaluated at the percolation temperature $T_p$, which is approximately equal to the temperature at the time of the bubble
nucleation, $T_p\approx T_n$, unless the phase transition occurs in a strongly supercooled state with $\alpha\gg 1$~\cite{Ellis:2019oqb,Ellis:2020nnr}.
In this work, we will confine ourselves to $\alpha<1$ and use the approximation $T_*\approx T_n$ to describe the GW profiles. 


\section{GW profiles}\label{sec:GW}

In general, there are three types of phase transitions: (i) non-runaway phase transition in a plasma, (ii) runaway phase transition in a plasma, and  (iii) runaway phase transition in vacuum. The last situation generically predicts  $\alpha\gg 1$ and hence exhibits a supercooled regime. When the supercooling occurs, one is further required  to check if the phase transition can   successfully complete, which is not automatically satisfied in a given model.   For the concern in this work, we will consider   $\alpha<1$. To further specify if the dark  phase transition is a   runaway or non-runaway type,  one may compare $\alpha$ with the threshold value $\alpha_\infty$~\cite{Espinosa:2010hh}\footnote{In Eq.~\eqref{eq:alphainfty},
	we have   taken only the dominant contribution from the gauge boson.}: 
\begin{align}\label{eq:alphainfty}
	\alpha_\infty\approx \frac{90g^2}{24\pi^2 g_\rho(T_n)}\left(\frac{\varphi_{\rm true}(T_n)}{T_n}\right)^2,
\end{align}
at which the wall of the scalar-field bubble begins to run away at the speed of light $v_w=1$. Usually it is thought that runaway phase transition occurs for $\alpha>\alpha_\infty$. Nevertheless, it is recently found that the efficiency of the vacuum energy conversion into the scalar field is suppressed by additional Lorentz factor ratios~\cite{Hoche:2020ysm,Ellis:2020nnr},\footnote{See, however, Refs.~\cite{Azatov:2020ufh,Gouttenoire:2021kjv} for  different perspectives.} so that the  efficiency is smaller than earlier estimates.  It leads to the situation where most   phase transitions originally found as a runaway type  now become a non-runaway type unless  the phase transitions are strongly supercooled, as confirmed in Refs.~\cite{Alanne:2019bsm,Schmitz:2020syl}.

In this paper, we will therefore consider the non-runaway phase transition with $\alpha<1$.  For a non-runaway phase transition, the dominant source to generate the stochastic GW  comes from the sound wave~\cite{Caprini:2015zlo}\footnote{
	We will not include the magneto-hydrodynamic turbulence in this work, but the conclusion drawn in this section is generic for an efficiency factor $\epsilon<1$ that determines how much portion of the energy from the  sound wave is turbulent~\cite{Caprini:2009yp}.}.
The induced GW
 peak amplitude and peak frequency are semi-analytically fitted  as~\cite{Caprini:2015zlo}
\begin{align}\label{eq:Omega_sw_p}
		\Omega_{\rm sw}^{\rm peak}h^2&\approx 2.65\times 10^{-6} (\mathcal{H} \tau_{\rm sw})\left(\frac{v_w}{\beta/\mathcal{H}}\right)\left(\frac{100}{g_\rho(T_n)}\right)^{1/3}\left(\frac{\kappa_{\rm sw} \alpha}{1+\alpha}\right)^2,
		\\[0.2cm]
	f_{\rm sw}^{\rm peak}&\approx 1.9\times 10^{-5}\,\text{Hz}\left(\frac{g_\rho(T_n)}{100}\right)^{1/6}\left(\frac{T_n}{100~\rm GeV}\right)\left(\frac{\beta/\mathcal{H}}{v_w}\right),\label{eq:f_sw_p}
\end{align}
where we have taken into account the suppression factor $\mathcal{H} \tau_{\rm sw}=\min(1,\mathcal{H}R_*/\bar{U}_f)$~\cite{Ellis:2018mja,Guo:2020grp,Ellis:2020awk} due to the finite lifetime of the sound wave,\footnote{The suppression factor is also given as $\Upsilon(\tau_{\rm sw})=1-(1+2\mathcal{H}\tau_{\rm sw})^{-1/2}$. For $\mathcal{H} \tau_{\rm sw} <1$ found in our numerical analysis, we arrive at $\Upsilon(\tau_{\rm sw})\approx  \mathcal{H}\tau_{\rm sw}$. } with the mean bubble separation $R_*=v_w (8\pi)^{1/3}/\beta$~\cite{Caprini:2019pxz} and $\bar{U}_f^2\approx 3\kappa_{\rm sw} \alpha/[4(1+\alpha)]$~\cite{Hindmarsh:2017gnf}. The  efficiency factor $\kappa_{\rm sw}$ represents the fraction  of the
released vacuum energy   converted into the bulk motion of the fluid.  For a  non-runaway phase transition, we apply the efficiency factor~\cite{Espinosa:2010hh}
\begin{align}
	\kappa_{\rm sw}\approx \frac{\alpha}{0.73+0.083\sqrt{\alpha}+\alpha}\,,
\end{align}
together with a   wall velocity $v_{w}=0.9$. After redshifting from the production at $T_n$, the GW spectrum $\Omega_{\rm GW}h^2\approx \Omega_{\rm sw}h^2$ today is given by
\begin{align}
	\Omega_{\rm sw}h^2(f)=	\Omega_{\rm sw}^{\rm peak}h^2 \mathcal{S}_{\rm sw}(f)\,,
\end{align}
where  $\Omega_{\rm sw}^{\rm peak}h^2$  is given by Eq.~\eqref{eq:Omega_sw_p} and the spectral shape function $\mathcal{S}_{\rm sw}$ in terms of the frequency $f$ reads
\begin{align}
\mathcal{S}_{\rm sw}=\left(\frac{f}{f_{\rm sw}^{\rm peak}}\right)^3\left[\frac{7}{4+3(f/f_{\rm sw}^{\rm peak})^2}\right]^{7/2},
\end{align}
with $f_{\rm sw}^{\rm peak}$ given by Eq.~\eqref{eq:f_sw_p}.

\begin{figure}[t]
	\centering
	\includegraphics[scale=0.9]{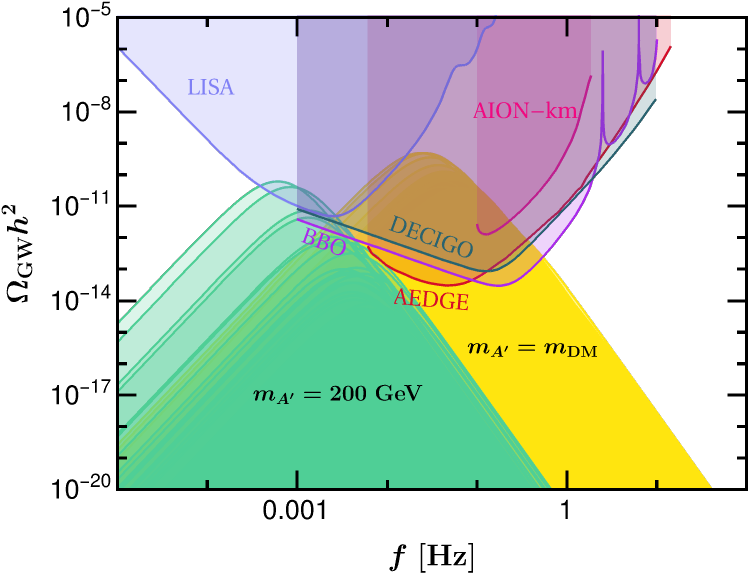}  
	\caption{The GW signals induced from the dark phase transition in the minimal boson plasma under the one-temperature treatment. Also shown are the sensitivity curves from space-based interferometers,  LISA~\cite{Caprini:2019egz}, BBO~\cite{Cutler:2005qq}, and  DECIGO~\cite{Seto:2001qf}, as well as  from   atomic interferometers, AEDGE~\cite{AEDGE:2019nxb} and AION~\cite{Badurina:2019hst} from km-scale terrestrial detectors. 
		\label{fig:GW_curves}
	}
\end{figure}

To obtain the GW curves, we perform  a numerical scan of the free parameters $\lambda, g_X$ in the following range\footnote{Using the analytic results shown in Appendix~\ref{appen:analytic_PT}, we find that the parameter space with  $\lambda \gtrsim 0.03$ and $\lambda\lesssim10^{-4}$ cannot generate strong enough GW signals, which is    confirmed in the numerical analysis.}:
\begin{align}\label{eq:scan}
	\lambda=5
	\times [10^{-4},10^{-2}]\,, \quad g_X=[0.1,1]\,.
\end{align}
The resulting $\Omega_{\rm sw}h^2$ is shown in Fig.~\ref{fig:GW_curves}. The shaded yellow region denotes the predicted GW signals under the scan~\eqref{eq:scan}, where the DM relic density is matched to the observed value via  Eq.~\eqref{eq:mAofg}. The peak frequencies for strong GW signals above $10^{-14}$ are located at $0.01-1$~Hz. In particular,   the GW curves have some overlap with the updated LISA sensitivity~\cite{Caprini:2019egz} only for   large $g_X\sim 1$. This result can also be inferred from the analytic phase transition presented in Appendix~\ref{appen:analytic_PT}. As indicated by Eq.~\eqref{eq:Omega_sw_p}, a large GW amplitude favors a small $\beta/\mathcal{H}$ and a large $\alpha$. From the left panel of Fig.~\ref{fig:analytic},  we see that a small $\beta/\mathcal{H}$ and a large $\alpha$ requires a large $g_X$. It implies that, when $A'_\mu$ is the DM with relic density fixed via Eq.~\eqref{eq:mAofg}, the dark vacuum $v_\phi$ is at $1-10$~TeV.
In general, for a non-supercooled phase transition,  we expect a nucleation temperature with the magnitude  not far from the dark vacuum $v_\phi$. Therefore, a large $\Omega_{\rm sw}h^2$ favors  $T_n> 100$~GeV. Then, from Eq.~\eqref{eq:f_sw_p}, we  see  that the peak frequency can be at $\mathcal{O}(1)$~mHz if $\beta/\mathcal{H}\lesssim 10$\footnote{For $\beta/\mathcal{H}\lesssim 10$ and $0.1<\alpha<1$, the dark phase transition might lead to the formation of primordial black holes~\cite{Liu:2021svg,Hashino:2022tcs}. We leave this interesting possibility for future studies.}. However,  as inferred from the left panel of Fig.~\ref{fig:analytic}, $ \beta/\mathcal{H}<10$ is hard to obtain, unless  $g_X\gtrsim 1$.  We then conclude that the inferred peak frequency for $g_X<1$  is above $\mathcal{O}(1)$~mHz  where the updated LISA sensitivity curve allows a GW amplitude as low as $\mathcal{O}(10^{-12})$. Therefore, only for some large $g_X$ will the induced GW curves have some overlap with the detectable region from  LISA.

It becomes clear that when $m_{A'}$ and $g_X$ are disentangled by  making the dark gauge boson unstable, the induced GW peak frequency can shift towards 1~mHz. Indeed, taking $m_{A'}=200$~GeV as an example, we can see from the green region that a relatively light dark gauge boson  allows more overlap with the detectable region from  LISA.

When $A'_\mu$ is the DM candidate within the minimal dark boson plasma, the strong first-order phase transition can readily induce GW signals that can be detected by the space-based interferometers DECIGO~\cite{Seto:2001qf} and BBO~\cite{Cutler:2005qq}. Furthermore, some proposed atomic interferometers can also capture the induced GW signals. This is particularly the case for the AION experiment with  km-scale terrestrial detectors~\cite{Badurina:2019hst}. The space experiment AEDGE~\cite{AEDGE:2019nxb} that uses cold atoms to search for ultra-light DM and GW is also able to capture the GW induced from the minimal dark bosonic phase transition. 
 
 \begin{table}
 	\centering
 	\renewcommand{\arraystretch}{1.4}
 	\begin{tabular}{l|c|c|c|c|c}
 		\hline\hline
 		BP            &  $\lambda$                    & $g_X$  & $T_n$    & $\alpha$     &  $\beta/\mathcal{H}$     \\ 
 		\hline 
 		LISA        &  $1.7\times 10^{-2}$     &  0.97   & 2.4~TeV   &  0.74           &  45             \\
 		\hline	
 		AION       &   $3.0\times 10^{-3}$    & 0.61    & 0.93~TeV &  0.80          &  417         \\
 		\hline
 		DECIGO    & $8.7\times 10^{-4} $     & 0.44   &  0.48~TeV &  0.89          &  1304           \\
 		\hline
 		BBO          &  $8.3\times 10^{-4}$     & 0.43   & 0.58~TeV   &  0.30        &    1561            \\
 		\hline	\hline
 	\end{tabular}
 	\caption{Selected BP for GW signals detectable by LISA, AION, DECIGO and BBO, respectively. }
 	\label{tab:4BP}
 \end{table}
 
Finally, we would like to comment on the LISA sensitivity curve adopted in this paper. In fact, there are some analyses of difference LISA sensitivity curves used in the literature. In particular, the analysis based on the peak-integrated-sensitivity curves in Ref.~\cite{Schmitz:2020syl} pointed out a lower curve of the LISA sensitivity. In this case,   the induced GW signals from the minimal dark boson plasma would have more overlap with  the estimated sensitivity region. In this paper, we do not show the    peak-integrated-sensitivity curve of LISA,  so our conclusions are based on the updated LISA sensitivity from Ref.~\cite{Caprini:2019egz}.

\section{Hear and see the dark: complementary tests of DM, GW and colliders}\label{sec:DM&GW}
 Given the current constraints and the future sensitivities, it is tempting to ask whether the parameter space for  a dark first-order phase transition, observable GW signals and DM favors a Higgs portal coupling that can be probed by  collider  detection. To this end, we select four BP from Fig.~\ref{fig:GW_curves}, which respectively corresponds to a GW curve overlapping with the LISA region (BP for LISA), residing between LISA and AION (BP for AION),   between AION and DECIGO (BP for DECIGO), and   between DECIGO and BBO (BP for BBO).  The parameters for each BP are shown in Tab.~\ref{tab:4BP}. 
 
 The first thing to notice from these BP is that, the predicted scalar masses are all above $2m_h$, so the constraints and future detection from the Higgs invisible decay are not applicable. In fact, a heavy dark scalar is generically expected when $A'_\mu$ is the DM candidate. As discussed in Sec.~\ref{sec:GW}, $g_X>0.1$ and $\lambda>10^{-4}$ are required to induce strong GW signals. Then we can infer from Eq.~\eqref{eq:mAofg} and Eq.~\eqref{eq: tree-level_mass} that the dark scalar mass is well above 10~GeV\footnote{It justifies the approximation of neglecting the dependence on the  recoil energy in the calculation of the DM-nucleon cross section.}. When $A'_\mu$ is no longer a stable DM, the dark scalar can be lighter such that the constraints and future detection of the Higgs invisible decay   come into play, as discussed in Ref.~\cite{Li:2023bxy}.

 \begin{figure}[t]
 	\centering
 	\includegraphics[scale=0.52]{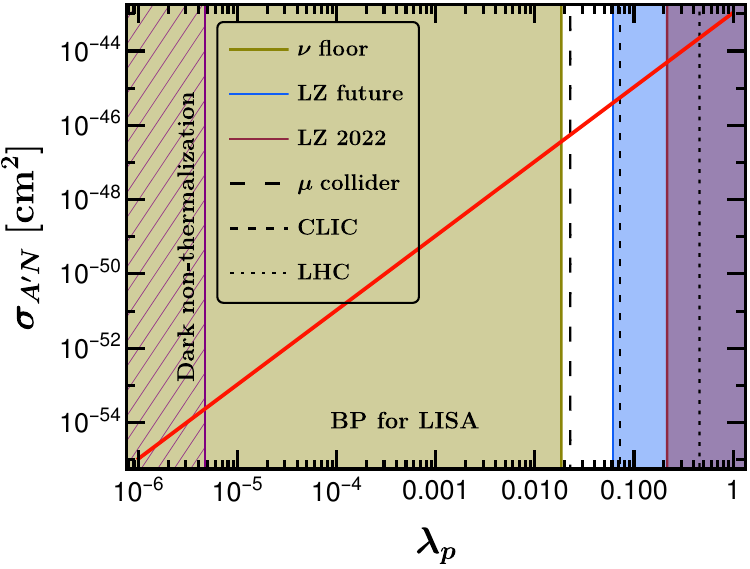}  \quad 
 	\includegraphics[scale=0.52]{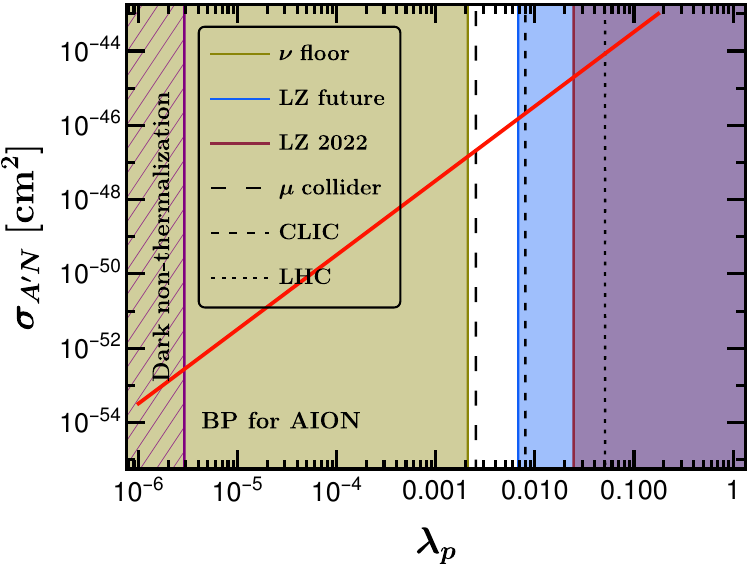} \\
 	\includegraphics[scale=0.535]{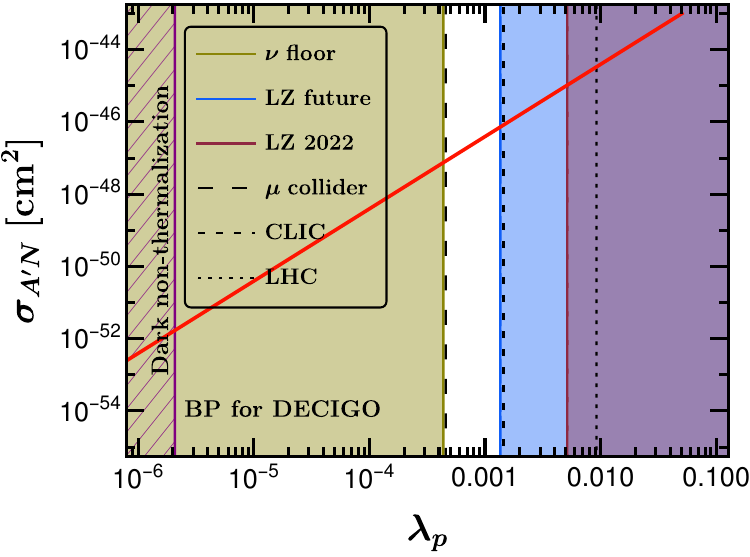} \quad
 	\includegraphics[scale=0.53]{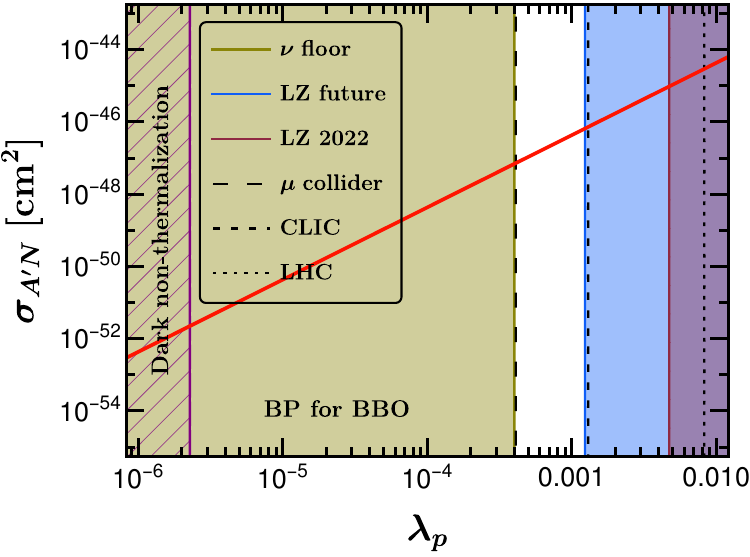}  
 	\caption{The parameter space for complementary tests from colliders, DM direct detection and GW interferometers. The spin-independent DM-nucleon cross section in terms of the Higgs portal coupling $\lambda_p$ is shown for each BP.
 		\label{fig:comp_4BP}
 	}
 \end{figure}

  For heavy dark scalar above $2m_h$, the  channel $\text{SM}\to \phi\to \text{SM}$  will serve as the direct collider probe. Currently, the constraints in    this channel are set by the LHC with an upper limit of $\theta$ at the level of 0.1. Future sensitivities from 3~TeV CLIC with an integrated luminosity $3~\text{ab}^{-1}$ in the decay channel $\phi\to 2h$ can reach a level of $\sin\theta^2\approx 0.001$~\cite{Linssen:2012hp,Buttazzo:2018qqp} and the 14~TeV muon
 collider with an integrated luminosity $14~\text{ab}^{-1}$ can further probe the mixing angle down to  $\sin\theta^2\approx 10^{-4}$ for $m_\phi<3$~TeV~\cite{Buttazzo:2018qqp}. The dashed lines shown in Fig.~\ref{fig:comp_4BP} (from right to left) correspond to the limit of $\theta$ excluded by current LHC, detectable by future CLIC and muon collider, respectively.   Note that, the indirect collider probes via the coupling strength modifier $\kappa_Z$ will reach a sensitivity of $\theta\simeq 0.06$, e.g., at the  ILC, CEPC and FCC-ee, which is comparable with the limit reachable by direct detection from the future 3~TeV CLIC.

 We show in Fig.~\ref{fig:comp_4BP} the red curve of the predicted spin-independent DM-nucleon cross section   in terms of the portal coupling in each BP. The rightmost shaded region in each panel  is excluded by the  LZ experiment~\cite{LZ:2022ufs}\footnote{Currently, the LZ experiment sets the most severe bound on the spin-independent DM-nucleon  cross section. See also   the very recently released constraint from the XENONnT experiment~\cite{XENONCollaboration:2023orw}.}. The  middle shaded region corresponds to the   future LZ sensitivity~\cite{LZ:2018qzl}. The leftmost shaded region corresponds to  the neutrino floor for xenon experiment~\cite{Ruppin:2014bra}, when the direct detection of DM suffers from irreducible  background of coherent neutrino-nucleus scattering.  We can see that the Higgs portal coupling  down to the level of $10^{-3}$ can be probed by the future LZ experiment, and  the spin-independent DM-nucleon cross section  can reach as large as $\sigma_{A'N}\simeq \mathcal{O}(10^{-45})~\text{cm}^2$.
 
 Combining the GW signals, DM direct detection and the collider tests,  we can see  that for all the BP, the CLIC is able to test a portion of the parameter space that induces GW signals observable by LISA, AION, DECIGO and BBO, respectively.  For all the panels shown in Fig.~\ref{fig:comp_4BP}, we can   see that the future muon collider  can probe a portal coupling  above the neutrino floor for the xenon experiment, and can fully test the parameter space  that predicts a spin-independent DM-nucleon cross section within the  future LZ sensitivity.  We then obtain an interesting situation for the minimal dark boson world:  when the stochastic GW signals are observed by future space-based/terrestrial interferometers, complementary probes via direct detection of a vector DM particle  and collider searches for a dark scalar   can provide cross-checks for the  microscopic  origin of   GW signals. Nevertheless, the indirect DM signals discussed in Sec.~\ref{sec:DM-indirect} are much suppressed. This comes from the fact that $m_{A'}\simeq\mathcal{O}(10)m_\phi$ is generally required when observational GW signals are generated during the dark phase transition. Then,  Eq.~\eqref{eq:mass-ratio}  is not   satisfied, as can also be seen from Tab.~\ref{tab:4BP} by using Eq.~\eqref{eq:coupling-ratio}. Besides, the parameter space shown in Fig.~\ref{fig:comp_4BP} and Tab.~\ref{tab:4BP} indicates that $\lambda_p\ll g_X^2$ and the DM mass is typically at TeV scale, making  $	(\sigma v)_{A'\to t}$ estimated in Eq.~\eqref{eq:A-SM}  several orders of magnitude smaller than the current bounds~\cite{Slatyer:2015jla,Planck:2018vyg}.

As mentioned in Sec.~\ref{sec:thermalization}, the dark thermalization conditions for the one-temperature treatment of the dark phase transition and DM freeze-out typically require  a Higgs portal coupling at $\lambda_{p}>\mathcal{O}(10^{-6})$. This is shown in Fig.~\ref{fig:comp_4BP}, where the purple meshed region denotes a small $\lambda_{p}$ that cannot establish a thermal contact between the SM and the dark sectors all the way down to the DM freeze-out temperature $T_{\rm fo}\simeq m_{A'}/25$. It points out that, for a Higgs-portal connected minimal dark sector,   the interested parameter space for strong GW signals and detectable DM indeed corresponds to a dark boson plasma sharing the same temperature with the SM  during the phase transition and the DM freeze-out.

Finally, it is worth mentioning that, while we focus on the one-temperature regime between the SM and dark sectors, the Higgs portal could   be too small to establish a common temperature, and the dark sector develops its own temperature with the initial condition  dating back to the post-inflationary epoch.  However, a much lower dark temperature will dramatically reduce the GW signals, while a higher dark temperature could be favored to boost the GW signals. In the latter case, the entropy injection to the SM plasma from the heavy dark sector would dilute the GW signals after phase transition. The net effect on  GW  signals  was studied in Ref.~\cite{Ertas:2021xeh}, pointing out that the enhancement effect due a higher dark temperature can dominate in a large region of parameter space. 
	 Different from the one-temperature scenario, where $\lambda_p$ shown in Fig.~\ref{fig:comp_4BP} indicates   scalar decay well before the  epoch of big-bang nucleosynthesis, indirect detection  for  a hotter dark sector with a much smaller  $\lambda_p$ could still be feasible in cosmology. Besides, the long-lived scalar could also feature  displaced vertices when produced at colliders. Since all these phenomena depend on 
 the parameter set $\{\lambda_{p},\lambda,m_{\phi}\}$,  one can expect   certain correlation  among GW, DM and indirect detection of the long-lived scalar   in cosmology and  colliders.


\section{Conclusion}\label{sec:cons}
We have presented a   dark boson plasma that can undergo a strong first-order phase transition, generate observable GW signals and provide a  DM candidate. The dark gauge boson, as the WIMP-like DM, plays a significant role in triggering the dark phase transition via finite-temperature effects and hence  inducing stochastic GW signals.   The theoretical framework is so simple and minimal that certain correlations exist in the complementary tests from GW  interferometers,  direct detection of DM and   colliders, all of which serve as promising avenues to help  us \textit{hear} and \textit{see} the dark. In particular, we found that the future muon collider in the decay channel $\phi\to 2h$ is able to probe the full parameter space where GW signals induced from the dark phase transition are detectable by LISA, DECIGO, BBO, AION and AEDGE,  and the spin-independent DM-nucleon cross section is reachable by the future LZ experiment above the neutrino floor.

\section*{Acknowledgements}
We would like to thank Michael J.~Ramsey-Musolf, Xun-Jie Xu and   Ke-Pan Xie for helpful discussions. S.-P.~Li  is supported in part by the National Natural Science Foundation of China under grant No.~12141501 and also by CAS Project for Young Scientists in Basic Research (YSBR-099). S.~K.~is supported in part  by JSPS KAKENHI Nos.~20H00160 and 23K17691.

\appendix
\section{An analytic perspective of the minimal dark phase transition}\label{appen:analytic_PT}
An analytic analysis of the dark phase transition allows us to get an efficient overview of the model parameter space where strong GW signals are favored and the  connection to DM could become clearer. In this appendix, we discuss the analytic dynamics of the dark phase transition by making some reasonable approximations. 
We will see that, such an analytic perspective is qualitatively  consistent with the numerical analysis performed in the main text.

One of the  widely considered potentials that allow  analytic treatments~\cite{Dine:1992wr,Quiros:1999jp}  has the following   polynomial  structure~\cite{Quiros:1999jp},
\begin{align}\label{eq:polypotential}
	V_{\rm poly}(\varphi,T)=D(T^2-T_0^2)\varphi^2-E T\varphi^3+\frac{\lambda}{4}\varphi^4\,,
\end{align} 
with field-independent constants $D,T_0, E$.
This polynomial potential is also   expected from the minimal dark $U(1)_X$ scenario considered in this work. Explicitly, it can be derived by using the high-temperature expansion of Eq.~\eqref{eq:JBint}~\cite{Dolan:1973qd,Laine:2016hma}, 
\begin{align}
	V_{T} & \approx \frac{T^{2}}{24} \sum_{i=\phi,\chi,A'} n_i m_{i}^{2}-\frac{T}{12\pi} \sum_{i=\phi,\chi,A'}n_i (m_{i}^{2})^{3/2}-\frac{1}{64\pi^{2}}\sum_{i=\phi,\chi,A'} n_i m_{i}^{4}\ln\left(\frac{m_{i}^{2}}{aT^{2}}\right),\label{eq:VT-2}
\end{align}
where $\ln a\approx5.41$  and we have dropped the $\varphi$-independent constants.  

The   $\mu$-dependence from $V_{\rm CW}$ in Eq.~\eqref{eq:VCW} can be removed by invoking the RG running of the tree-level parameters in $V_0$~\cite{Kastening:1991gv,Bando:1992np,Ford:1992mv}, with

\begin{align}\label{eq:running-lambda}
	\lambda(\mu)  &=\lambda (\mu_0)+\frac{3g_{X}^{4}}{16\pi^{2}}\ln\left(\frac{\mu^{2}}{\mu_{0}^{2}}\right)+\mathcal{O}(\lambda^{2})\,,
	\\[0.2cm]
	\mu_S^{2}(\mu)  &=\mu_S^{2}(\mu_0)+\frac{\lambda(\mu_0)}{4\pi^{2}}\mu_S^{2}(\mu_0)\ln\left(\frac{\mu^{2}}{\mu_{0}^{2}}\right).\label{eq:running-M}
\end{align}
After applying the daisy resummation~\eqref{eq:AEresum}, we finally arrive at the effective potential
\begin{align}\label{eq:effpotential}
	V_{{\rm eff}}(\varphi,T)  =\left(\frac{g_X^{2}}{8}T^{2}-\frac{\lambda v^2_\phi}{2}\right)\varphi^{2}-\frac{g_X^3T}{12\pi}\left[2\varphi^3+\left(\varphi^2+\frac{T^2}{3}\right)^{3/2}\right]+\frac{\lambda}{4}\varphi^{4}\,,
\end{align}
where only the dominant contribution  from the gauge boson is taken into account, and  corrections at $\mathcal{O}(\lambda^2)$ and terms proportional to $3g_X^4/(4\pi)^2$ in Eqs.~\eqref{eq:running-lambda} and~\eqref{eq:running-M} are neglected.\footnote{
	It can also be interpreted by taking $\lambda$ and $\mu_S^2$, as well as $g_X$ in Eq.~\eqref{eq:effpotential} evaluated at a fixed  RG scale $\mu=\mu_0$.} 

It can now be seen that, without the daisy resummation in Eq.~\eqref{eq:effpotential}, the effective potential reduces to the polynomial form~\eqref{eq:polypotential}, with 
\begin{align}\label{eq:DTE}
	D=\frac{g_X^2}{8}\,,\quad T_0=\frac{2\sqrt{\lambda}v_\phi}{g_X}\,, \quad E=\frac{g_X^2}{4\pi}\,.
\end{align}

The potential~\eqref{eq:polypotential} at high temperatures could  develop an energy barrier between the false and true vacua.
As usual,  there is a trivial solution ${\varphi}=0$ to $\partial  {V}_{\rm poly}/\partial  {\varphi}=0$,   corresponding to a local minimum if $4\lambda v_\phi^2<g_X^2  {T}^2$. There are two    additional  solutions to $\partial {V}_{\rm poly}/\partial {\varphi}=0$, with
\begin{align}\label{eq:nontrival-extrema}
	{\varphi}_{\pm}= \frac{1}{8\pi\lambda }\left[3g_X^3  T \pm(9g_X^6{T}^2-16\pi^2 g_X^2 \lambda {T}^2+64\pi^2\lambda^2v_\phi^2)^{1/2}\right].
\end{align}
To make these extrema physical, we should set 
\begin{align}\label{eq:analytic_con}
	64\pi^2\lambda^2 v_\phi^2-16\pi^2 g_X^2 \lambda T^2+9g_X^6 T^2>0\,,
\end{align}
for temperatures $T$ during the phase transition. 
The local maximum at  ${\varphi}_-$ is expected to develop an energy barrier between the false vacuum at ${\varphi}=0$ and the true vacuum at ${\varphi}_{+}$.

\begin{figure}[t]
	\centering
	\includegraphics[scale=0.55]{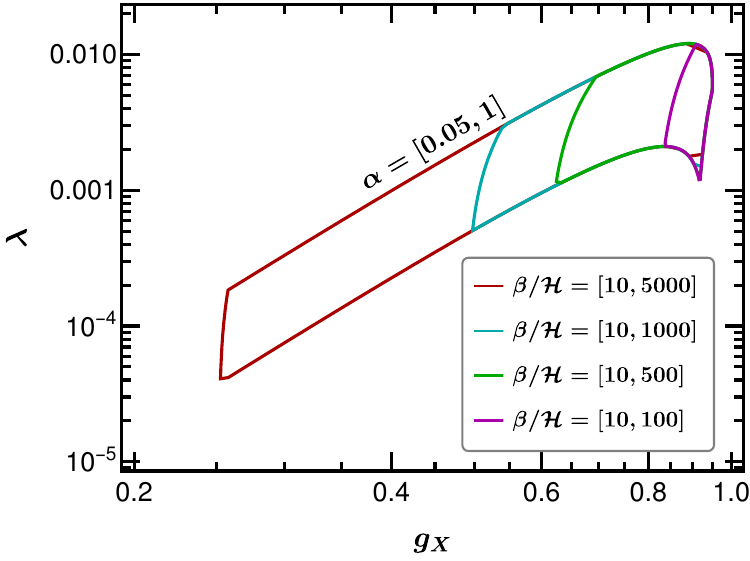} \quad
		\includegraphics[scale=0.45]{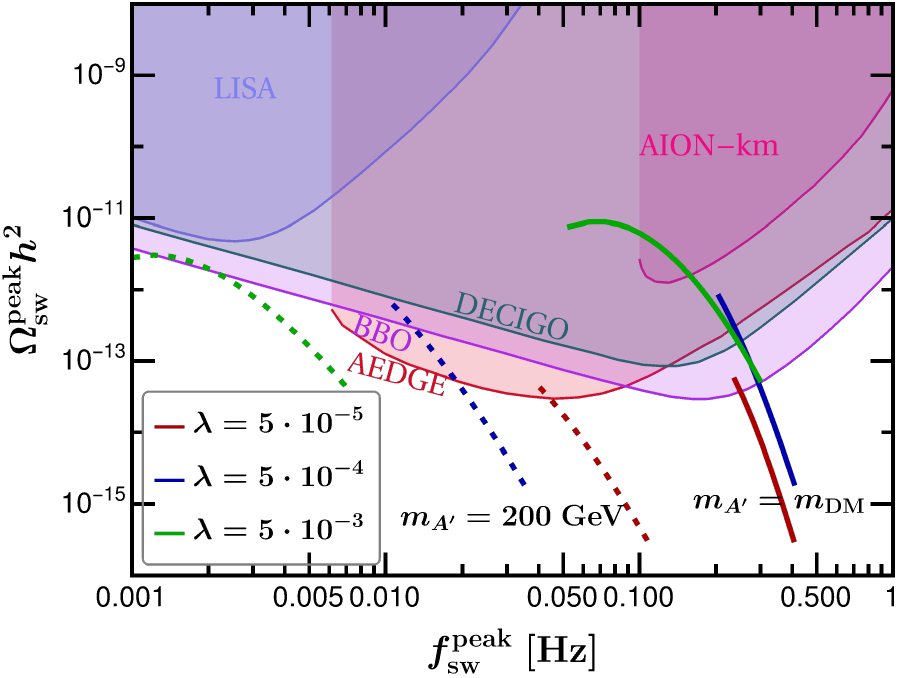}
	\caption{Left: the parameter space in terms of $\lambda$ and $g_X$ for $0.05<\alpha<1$ and for some typical ranges of $\beta/\mathcal{H}$. Right:  the correlation between the GW peak amplitude $\Omega_{\rm sw}^{\rm peak}h^2$ and peak frequency $f_{\rm sw}^{\rm peak}$ for some benchmark values of $\lambda$ under the scan of $0.1<g_X<1$. The solid lines are obtained via Eq.~\eqref{eq:mAofg} while the dotted lines are obtained by fixing $m_{A'}=200$~GeV without assuming a DM candidate. 
			\label{fig:analytic}
		}
\end{figure}

The Euclidean action $S_3$ from the polynomial potential has the following simple analytic solution~\cite{Adams:1993zs},   
\begin{align}
	S_3\approx \frac{64\pi ET \sqrt{\delta}(\beta_1\delta+\beta_2\delta^2+\beta_3\delta^3)} {81\lambda^{3/2}(2-\delta)^2}\,,
\end{align}
where $\delta\equiv 2\lambda D(1-T_0^2/T^2)/E^2$, with $D,T_0$ and $E$ given in Eq.~\eqref{eq:DTE}, and $\beta_1=8.2938,\beta_2=-5.5330,\beta_3=0.8180$. Using this analytic expression, we can evaluate the nucleation temperature via Eq.~\eqref{eq:S3overT-1}, and subsequently the $\alpha,\beta/\mathcal{H}$ parameters. 

We show in the left panel of Fig.~\ref{fig:analytic} the parameter space of $\lambda, g_X$ for $0.05<\alpha<1$ and some typical ranges of $\beta/\mathcal{H}$.
 By taking some BP of $\lambda$ inferred from the left panel, we scan the gauge coupling in the range: $[0.1,1]$, so as to obtain the relation between $f_{\rm sw}^{\rm peak}$ and $\Omega_{\rm sw}^{\rm peak}h^2$ from Eqs.~\eqref{eq:Omega_sw_p} and~\eqref{eq:f_sw_p}, as shown  in the right panel of  Fig.~\ref{fig:analytic}.   Both the short solid and dotted lines are cut by the conditions $0.1<\alpha<1, \beta/\mathcal{H}>1$ and~\eqref{eq:analytic_con}.

It can be seen that, when $A'_\mu$ is the DM candidate, the condition from Eq.~\eqref{eq:mAofg} implies a peak frequency at sub-Hz, with the peak amplitude detectable by AION, AEDGE, DECIGO and BBO. When relaxing the DM condition, we take a lower mass for the dark gauge boson, $m_{A'}=200$~GeV, which shifts the peak frequency towards mHz. Note that the LISA sensitivity curve is taken from the analysis given in Ref.~\cite{Caprini:2019egz}, rather than the peak-integrated-sensitivity curves from Refs.~\cite{Alanne:2019bsm,Schmitz:2020syl}.

\section{DM annihilation rates}\label{appen:DM_rates}

The gauge-scalar interaction is derived from the gauge kinetic term:
\begin{align}
\mathcal{L}_{\rm kin}=(\partial_{\mu}-i g_X A'_{\mu}) S \cdot(\partial^{\mu}+i g_X A^{\prime \mu})S^*,
\end{align}
with $S=(v_\phi+\phi)/\sqrt{2}$ in the unitary gauge. It leads
to the Feynman  vertices,
\begin{align}
	A^{\prime \mu}A^{\prime \nu}\phi & :2g_X^{2}v_\phi \eta^{\mu\nu}\, ,\qquad A^{\prime \mu}A^{\prime \nu}\phi^{2}:2g_X^{2}\eta^{\mu\nu}\,,
\end{align}
where $\eta^{\mu\nu}$ is the Minkowski metric.
 For the dark scalar $\phi$, the Feynman vertices
from   potential~\eqref{eq:tree-level-potential} are given by
\begin{align}
		\phi^{3} & :-6\lambda v_{\phi}\, ,\qquad \phi^{4}:-6\lambda\,.
\end{align}

The dark gauge boson freeze-out is dominated by the quartic  contact interaction
and the $t,u$-channel annihilation, while the $s$-channel
annihilation carries an additional factor $\lambda^{2}$ and hence
is much smaller than the $t,u$ channels for  $\lambda \ll g_X^{2}$. Neglecting the $s$-channel is
a good approximation since there is no $s$-channel resonant production
of $\phi$ for  $m_{A'}\gg m_{\phi}$. 
The cross section of $2A'_\mu\to2\phi$ can be parameterized as
\begin{align}\label{eq:sigmav}
	\sigma v & \equiv 2(a+bv^{2})\, ,
\end{align}
where  the factor of 2 takes into account the pair DM annihilation,   and $v$ is the relative velocity of incoming DM particles in the lab frame.
To determine the cross section, let us now calculate the amplitude of $2A'_\mu\to 2\phi$, with the Feynman diagrams shown in Fig.~\ref{fig:2X2phi}. The amplitudes in the center-of-mass frame 
are given by
\begin{align}
	\mathcal{M}_{t}&=\frac{8g_X^4v_\phi^2\eta^{\mu\nu}\epsilon_{\mu}(p_{1})\epsilon_{\nu}(p_{2})}{m_\phi^2-\hat{s}(1-\beta_{A'}\beta_\phi\cos\theta)}\,,
	\\[0.2cm]
	\mathcal{M}_{u}&=\frac{8g_X^4v_\phi^2\eta^{\mu\nu}\epsilon_{\mu}(p_{1})\epsilon_{\nu}(p_{2})}{m_\phi^2-\hat{s}(1+\beta_{A'}\beta_\phi\cos\theta)}\,, \\[0.2cm]
	\mathcal{M}_{c}&=2g_X^{2}\eta^{\mu\nu}\epsilon_{\mu}(p_{1})\epsilon_{\nu}(p_{2})\,,
\end{align}
for the $t,u$ channels and  contact annihilation, respectively. Here $\hat s$ is the Lotentz-invariant Mandelstam variable, $\beta_{A',\phi}\equiv \sqrt{1-4m_{A',\phi}^2/\hat s}$, and $\theta$ is the angle between the momenta of the  incoming and outgoing particles in the center-of-mass frame. 


The differential cross section  in the center-of-mass frame yields
\begin{align}
		\frac{d\sigma}{d\Omega}=\frac{1}{S(2E_{1})(2E_{2})|\vec{v}_{1}-\vec{v}_{2}|}\left[\frac{1}{16\pi^{2}}\frac{|\vec{p}_{f}|}{E_{{\rm cm}}}|\mathcal{M}|^2_{\rm sum}\right]=\frac{\beta_\phi |\mathcal{M}|^2_{\rm sum}}{128\pi^2 \beta_{A'}\hat{s}}\,,
\end{align}
	where $\mathcal{M}_{\rm sum}=\mathcal{M}_t+\mathcal{M}_u+\mathcal{M}_c$,  $1/S=1/2$ is the symmetry factor accounting for the identical scalar
in the final states, and $E_{\rm cm}=\sqrt{\hat s}$.  The final-state momentum $|\vec{p}_{f}|$ in the center-of-mass frame can be expressed in terms of $\hat s$, with $|\vec{p}_{f}|=\beta_\phi\sqrt{\hat s}/2$. 

Going 
to the frame where one of the incoming particles is at rest (the lab frame), as defined in Eq.~\eqref{eq:sigmav},  we can parameterize
the incoming four-momenta  as $p_{1}=(E_{},\vec{p}),p_{2}=(m_{A'},0)$, with the relative velocity
$v=|\vec{p}/E|$. Given  $E_{}=m_{A'}/\sqrt{1-v^{2}}, |\vec p|=m_{A'} v/\sqrt{{1-v^{2}}}$,
$\hat s$  in the lab frame is given by
\begin{gather}
	\hat s=(p_{1}+p_{2})^{2}
	=2m_{A'}^{2}(1+\frac{1}{\sqrt{1-v^{2}}})\approx4m_{A'}^{2}(1+v^{2}/4)\,.
\end{gather}
Note that from $\hat{s}$ in the center-of-mass frame we obtain $v^2=v_{\rm cm}^2/2$, where $v_{\rm cm}$ is the relative velocity in the center-of-mass frame.
 It is now straightforward to evaluate the cross section in the lab frame. Expanding the small velocity up to $\mathcal{O}(v^2)$ and neglecting the scalar mass, we obtain the $a,b$ parameters as 
\begin{align}
	a  \approx \frac{3g_X^4}{16\pi m_{A'}^2}\,,\quad
	b \approx -\frac{15g_X^4}{128\pi m_{A'}^2}\,.
\end{align}

Based on  Eq.~\eqref{eq:sigmav},   the collision rate $\gamma_{A'}$ can be written as
\begin{align}\label{eq:collision_rate}
	\gamma_{A'} &=\int d\Pi_{1}d\Pi_{2} d\Pi_{3}d\Pi_{4}f_{1}(E_{1})f_{2}(E_{2}) |\mathcal{M}|_{\rm sum}^2 (2\pi)^4 \delta^4\left(p_1+p_2-p_3-p_4\right)
  \\[0.2cm]
	& =\int d\Pi_{1}d\Pi_{2}f_{1}(E_{1})f_{2}(E_{2})[4\sigma v E_{1}E_{2}]
 \\[0.2cm]
	&=2(a+6b/x) n_{A'}^2 \label{eq:a+3b/x}
\\[0.2cm]&	\equiv \langle \sigma v\rangle_{A'} n_{A'}^2\,,\label{eq:signav_def}
\end{align}
where  $d\Pi_{i}\equiv d^3p_i/[(2\pi)^32 E_i]$ is the phase-space factor with $E_i$ the energy,   $f_i(E_i)$ is the thermal distribution function of $A'_\mu$ with the number density $n_{A'}$, and the quantum statistics for the final states is neglected. Note that in  Eq.~\eqref{eq:a+3b/x},  $x\equiv m_{A'}/T$, and  the $b$-dependent term is   derived by taking the relative velocity $v^2=|\vec{p}_1/E_1-\vec{p}_2/E_2|^2\approx (p_1^2+p_2^2-2\vec{p}_1\cdot \vec{p}_2)/m_{A'}^2$ in the phase-space integration with the Boltzmann distribution $f_i(E_i)=e^{-E_i/T}$, without introducing the M\o ller velocity~\cite{Gondolo:1990dk}.

\bibliographystyle{JHEP}
\bibliography{Refs}

\end{document}